\documentclass{Interspeech}

\interspeechcameraready

\usepackage{setspace} %
\usepackage{footnotebackref}
\usepackage{booktabs} %
\usepackage{multirow} %
\usepackage{tabularray}
\usepackage[T1]{fontenc}
\usepackage[labelfont=bf]{caption} %
\usepackage{color}
\usepackage[table]{xcolor}
\usepackage{cite}
\usepackage{enumitem}
\usepackage{supertabular}
\usepackage{graphicx}

\hypersetup{
    colorlinks=true,
    linkcolor=blue,
    filecolor=magenta,      
    urlcolor=magenta,
    citecolor=magenta,
    pdftitle={Lessons Learned from the URGENT 2024 Speech Enhancement Challenge},
    pdfpagemode=FullScreen,
}

\usepackage{tikz}
\newcommand*\circled[1]{\tikz[baseline=(char.base)]{
            \node[shape=circle,draw,inner sep=0.8pt] (char) {#1};}}

\renewenvironment{thebibliography}[1]{
  \begin{oldthebibliography}{#1}
  \setlength{\itemsep}{-0.05em}
  \setlength{\parskip}{0.em}
}
{
  \end{oldthebibliography}
}

\makeatletter
\def\bstctlcite{\@ifnextchar[{\@bstctlcite}{\@bstctlcite[@auxout]}}
\def\@bstctlcite[#1]#2{\@bsphack
  \@for\@citeb:=#2\do{%
    \edef\@citeb{\expandafter\@firstofone\@citeb}%
    \if@filesw\immediate\write\csname #1\endcsname{\string\citation{\@citeb}}\fi}%
  \@esphack}
\makeatother

\title{Lessons Learned from the URGENT 2024 Speech Enhancement Challenge}

\author[affiliation={1}]{Wangyou}{Zhang}
\author[affiliation={2}]{Kohei}{Saijo}
\author[affiliation={3}]{Samuele}{Cornell}
\author[affiliation={4}]{Robin}{Scheibler}
\author[affiliation={1}]{Chenda}{Li}
\author[affiliation={5}]{Zhaoheng}{Ni}
\author[affiliation={5}]{Anurag}{Kumar}
\author[affiliation={6}]{Marvin}{Sach}
\author[affiliation={1}]{Wei}{Wang}
\author[affiliation={6}]{Yihui}{Fu}
\author[affiliation={3}]{Shinji}{Watanabe}
\author[affiliation={6}]{Tim}{Fingscheidt}
\author[affiliation={1}]{Yanmin}{Qian}

\affiliation{}{%
  Shanghai Jiao Tong University, China
  $^2$Waseda University, Japan
  $^3$Carnegie Mellon University, USA
  $^4$Google DeepMind, Japan
  $^5$Meta, USA
  $^6$Technische Universität Braunschweig}{Germany}
\email{wyz-97@sjtu.edu.cn,saijo@pcl.cs.waseda.ac.jp}


\usepackage{comment}

\begin{document}
\bstctlcite{IEEEexample:BSTcontrol} %

\maketitle

\begin{abstract}
    The URGENT 2024 Challenge aims to foster speech enhancement (SE) techniques with great universality, robustness, and generalizability,
    featuring a broader task definition, large-scale multi-domain data, and comprehensive evaluation metrics.
    Nourished by the challenge outcomes, this paper presents an in-depth analysis of two key, yet understudied, issues in SE system development: data cleaning and evaluation metrics.
    We highlight several overlooked problems in traditional SE pipelines:
    (1) mismatches between declared and effective audio bandwidths, along with label noise even in various ``high-quality'' speech corpora;
    (2) lack of both effective SE systems to conquer the hardest conditions (e.g., speech overlap, strong noise\,/\,reverberation) and reliable measure of speech sample difficulty;
    (3) importance of combining multifaceted metrics for a comprehensive evaluation correlating well with human judgment.
    We hope that this endeavor can inspire improved SE pipeline designs in the future.
\end{abstract}

\vspace{-10pt}
\section{Introduction}
\label{sec:intro}
\vspace{-4pt}
The URGENT 2024 Challenge~\cite{URGENT-Zhang2024} is a newly launched competition that aims to assess and advance the universality, robustness, and generalizability of speech enhancement (SE) techniques.
It features a broader SE task definition, large-scale and diverse source data, and extensive evaluation metrics, complementing existing SE challenges~\cite{ConferencingSpeech-Rao2021,ICASSP-Dubey2023,ICASSP-Ristea2024,INTERSPEECH2022-Diener2022,ICASSP23AEC-Cutler2023,2nd-Akeroyd2023,L3DAS23-Marinoni2024,CHiME_7-Leglaive2023}.
Specifically, this challenge encourages research towards a single universal SE system that can handle speech signals with various sampling frequencies (SF) corrupted by diverse distortions, including additive noise, reverberation, clipping, and bandwidth limitation.
To comprehensively assess SE performance, this challenge adopts as many as 13 evaluation metrics.
The ranking of different submissions is then determined using a dedicated algorithm inspired by the Friedman test~\cite{Use-Friedman1937} that balances the metric performances from different perspectives~\cite{URGENT-Zhang2024}.

The challenge consists of three phases, i.e., \emph{validation}, \emph{non-blind test}, and \emph{blind test} phases.
Each phase provides a corresponding dataset with 1000 samples for leaderboard evaluation.
In the first two phases, all samples are simulated using unseen speech and noise data, and the corresponding ground-truth clean speech signals (labels) are provided after each phase ends.
This allows for fast system development as all evaluations can be done automatically using objective metrics.
In the final phase, only half of the samples are simulated, while the other half are collected from real-world scenarios, allowing for a more realistic and comprehensive evaluation. 
The final ranking is fully determined by the results obtained in the final phase\footnote{\url{https://urgent-challenge.com/competitions/5}.}.
Through the above procedure, the challenge has received submissions from 21 participating teams in the final blind test phase.

Different from \cite{URGENT-Zhang2024} that only provides a basic introduction of the challenge, this paper aims to share the learnings from the challenge design choices as well as the final competition results.
Specifically, the goal of this paper is to uncover overlooked issues and highlight potential improvements in the design of data and evaluation metrics, two critical components of system development.
Below, we summarize our key findings.
\begin{itemize}[wide]
	\item[1)] A mismatch between declared and effective audio bandwidths commonly exist in both public audio corpora and in-the-wild data.
	For example, \textasciitilde{}25\% LibriTTS~\cite{LibriTTS-Zen2019}, \textasciitilde{}100\% DNS5 LibriVox~\cite{ICASSP-Dubey2023}, and \textasciitilde{}100\% CommonVoice 11.0~\cite{CommonVoice-Ardila2020} English speech data have mismatched audio bandwidths.
	Furthermore, the label noisiness issue is also surprisingly common, even in supposedly clean corpora such as WSJ~\cite{WSJ0-LDC1993,WSJ1-Consortium1994}, VCTK~\cite{VCTK-Veaux2013} and LibriTTS.
    Both issues can cause unexpected model behaviors and require extra care in data preparation.
	\item[2)] Overlapped speech, strong wideband\,/\,instantaneous noise, and high reverberation remain the hardest challenges in SE tasks. Moreover, there is no good objective measure of the sample difficulty for the SE task, and conventional metrics such as signal-to-noise ratio (SNR) are often misleading.
    \item[3)] Interestingly, some newly proposed non-intrusive metrics such as UTMOS~\cite{UTMOS-Saeki2022} and VQScore~\cite{VQScore-Fu2024} show better perceptual correlations than traditional ones (e.g., DNSMOS~\cite{DNSMOS-Reddy2022}), despite that they may not be designed directly for the SE task.
	\item[4)] It is beneficial to combine various metrics for a multifaceted evaluation (cf. Section~\ref{ssec:metric}) of the SE performance, which leads to high correlations with human judgments.
    On the other hand, using solely a \emph{single} category of metrics, especially non-intrusive metrics, for SE performance assessment can be unreliable or even misleading.
\end{itemize}

\begin{figure*}
  \centering
  \includegraphics[width=0.8\textwidth]{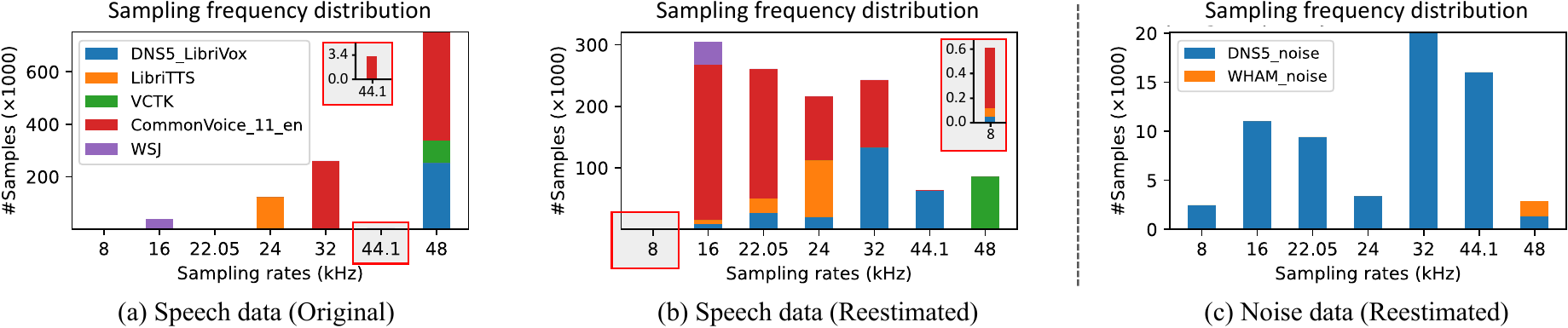}
  \caption{Sampling frequency distribution of speech and noise data in the training set. The original noise data are all in 48 kHz.}
  \label{fig:fs_dist}
  \vspace{-1em}
\end{figure*}
\begin{figure*}
  \centering
  \includegraphics[width=0.8\textwidth]{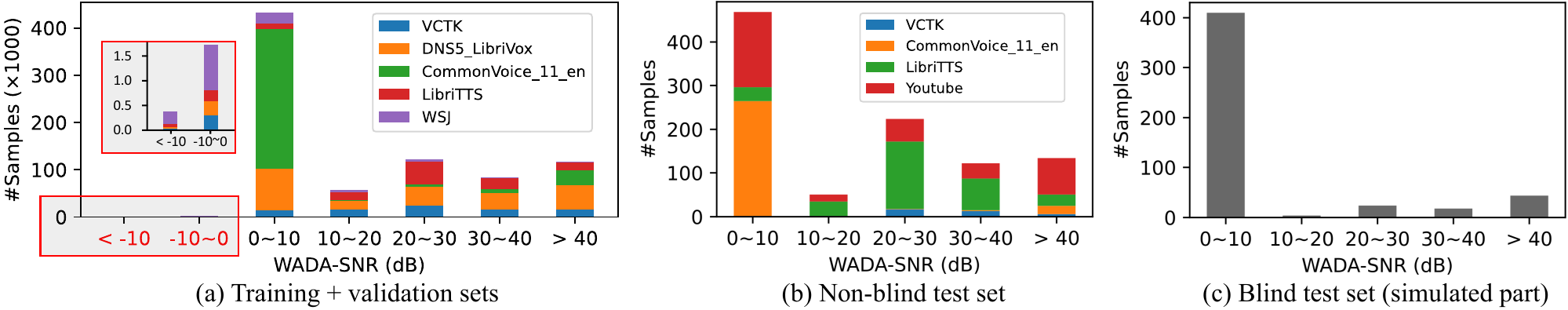}
  \caption{Histogram of estimated SNRs of the ``clean'' speech labels in each dataset based on the WADA algorithm~\cite{WADASNR-Kim2008}.}
  \label{fig:wada_snr}
  \vspace{-2em}
\end{figure*}

\vspace{-8pt}
\section{Data Analysis}
\label{sec:data}
\vspace{-3pt}

\subsection{Training and validation data}
\label{ssec:tr_val_data}

Both training and validation datasets of the URGENT 2024 Challenge are fully simulated based on 5 public speech corpora (DNS5 LibriVox speech~\cite{ICASSP-Dubey2023}, LibriTTS, CommonVoice 11.0 English portion~\cite{CommonVoice-Ardila2020}, VCTK, and WSJ),
2 noise corpora (Audioset+FreeSound noise in DNS5 challenge~\cite{ICASSP-Dubey2023}, and WHAM! noise recordings~\cite{WHAM-Wichern2019}),
and simulated room impulses (RIRs) from the DNS5 challenge data\footnote{Readers can refer to~\cite{URGENT-Zhang2024} for more details.}.
The curated dataset includes \textasciitilde{}1300 hours of source speech data and \textasciitilde{}250 hours of noise data.

When revisiting the preprocessing of training and validation data (i.e., filtering and resampling), we found two major obstacles that may lead to compromised data quality. While Zhang \textit{et al.}~\cite{URGENT-Zhang2024} managed to address the first one, the other remains an open question that requires more investigations in future work. Below we discuss each problem in detail.
\begin{itemize}[wide,labelwidth=!,labelindent=0pt]
	\item[1)] \textbf{Effective bandwidth vs. actual sampling frequency}: We found that various public speech corpora contain ``bad'' samples with a lower effective bandwidth than the SF's Nyquist frequency, suggesting that these samples have been resampled from a lower SF. 
	For example, some LibriTTS samples only contain frequency components up to 4 kHz while their SF is always 24 kHz, leading to wasted storage and potential misuses.
	This is particularly important in our challenge as precise information about the effective bandwidth is pivotal for unifying bandwidth extension (BWE) and other SE subtasks. It is also crucial
	 in speech generation tasks such as text-to-speech (TTS) and voice conversion (VC), where such ``bad'' samples may mislead the learning procedure.
	To address this issue, the challenge adopted an energy-based algorithm~\cite{Hi_Fi-Bakhturina2021} to estimate the effective bandwidth of all audio samples.
	In Fig.~\ref{fig:fs_dist}, we provide a breakdown of the SF distributions of different training corpora (speech and noise) before and after bandwidth re-estimation.
	Surprisingly, except for WSJ, VCTK, and WHAM!, other commonly-used corpora, i.e., DNS5, LibriTTS, and CommonVoice all suffer from the above issue, with \textasciitilde{}100\%, \textasciitilde{}25\%, and \textasciitilde{}100\% samples having mismatched audio bandwidths, respectively.
	This also calls for better speech data acquisition pipelines that can address this mismatched bandwidth issue.
	\item[2)] \textbf{Label noisiness}: Conventional SE models generally demand clean speech as training labels, i.e., without any noise.
        In practice, however, we found that various supposedly clean speech corpora (e.g., WSJ, VCTK, and LibriTTS) still contain low-quality speech samples with non-negligible noise.
	Even with multiple filtering strategies in~\cite{URGENT-Zhang2024}, e.g. voice activity detection (VAD) and DNSMOS-based filtering, the resultant speech data still inevitably contain noisy speech labels for training.
    This indicates the deficiency of existing speech data filtering strategies, which remain an open problem.

\noindent To better understand the SNR distribution of speech labels in the challenge datasets, we adopt the non-intrusive WADA algorithm~\cite{WADASNR-Kim2008} to roughly estimate sample-wise SNRs.
	The histogram of WADA-SNRs is depicted in Fig.~\ref{fig:wada_snr} (a).
	Although the absolute values can be inaccurate due to the limitation of the WADA algorithm, the relative range of WADA-SNRs can still provide a clue about the amount of label noise.
	For example, there are \textasciitilde{}2000 speech labels with negative WADA-SNRs, which we manually verified to be mostly noisy.
	Their source corpora (i.e., LibriTTS, VCTK, WSJ, etc.) are also widely used in the SE literature, which implies that \emph{most SE studies are implicitly conducting a sort of noisy-target training (NyTT)~\cite{NyTT-Fujimura2021,Analysis-Fujimura2023} instead of clean-target training}\footnote{Note that it may not be the authentic NyTT as the domains of the inherent label noise and additionally mixed noise are not explicitly taken care of to facilitate NyTT~\cite{Analysis-Fujimura2023}.}.

\noindent Furthermore, to validate the effectiveness of NyTT, we calculate the WADA-SNRs on the enhanced version of the ``clean'' speech labels based on the official baseline model, TF-GridNet~\cite{TF_GridNet-Wang2023}.
The corresponding histograms\footnote{Available at \url{https://github.com/urgent-challenge/urgent2024_analysis/blob/main/wada_snr/}.} turned out to be highly similar to Fig.~\ref{fig:wada_snr}, implying that the enhanced speech labels can still be noisy.
We also manually verified that some enhanced audios with low WADA-SNRs indeed contain noise of different levels.
This indicates that \emph{the SE model may be misguided to preserve the noise floor (usually at a low level) in the enhanced speech if the training labels are more or less noisy}.
This calls for more advanced techniques for leveraging partially noisy speech data for SE training. Addressing this challenge is one of the key objectives of the follow-up URGENT 2025 Challenge~\cite{URGENT2025}.
\end{itemize}

\begin{figure*}
  \centering
  \includegraphics[width=0.9\textwidth]{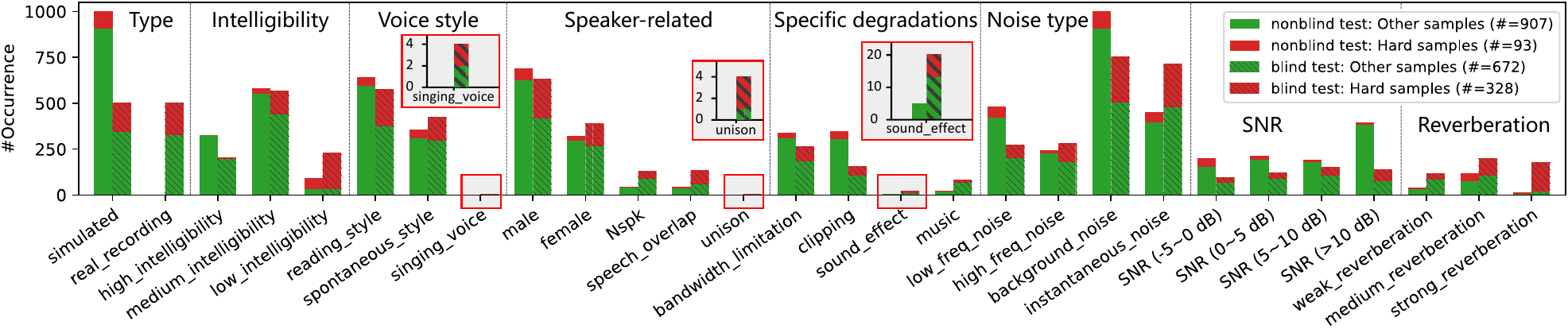}
  \caption{Tag occurrences of different samples in the non-blind and blind test sets. The ``hard samples'' are defined as samples with poor metric scores below pre-defined thresholds. The detailed rule for determining hard samples can be found in Section~\ref{ssec:nbt_data}.}
  \label{fig:hard_samples_test}
  \vspace{-1.5em}
\end{figure*}

\vspace{-8pt}
\subsection{Non-blind test data}
\label{ssec:nbt_data}
\vspace{-3pt}

The official non-blind test dataset consisting of $1000$ samples is carefully designed to contain unseen speech, noise, and real room impulse response (RIR) data from different domains.
The speech data include curated samples from \circled{\scriptsize 1} the test partitions of CommonVoice, LibriTTS, and VCTK, and \circled{\scriptsize 2} crawled Youtube videos (mostly podcasts with various topics) with Creative Commons licenses.
The former (totally 641 samples) accounts for matched domains with reading-style speech, while the latter (totally 359 samples) represents mismatched domains including both spontaneous and reading styles.
The noise data include curated samples from \circled{\scriptsize 1} the test partition of WHAM! and \circled{\scriptsize 2} two unseen noise corpora (i.e., DEMAND~\cite{DEMAND-Thiemann2013}, and TUT Urban Acoustic Scenes 2018 evaluation dataset~\cite{Multi_Device-Mesaros2018}) covering more scenes and noise types.
The SNR range is set to [-5, 20] dB for data simulation.
The RIR data consists of real RIRs curated from SLR28~\cite{SLR28-Ko2017} (real partition) and MYRiAD v2~\cite{MYRiAD-Dietzen2023}, covering a wide reverberation time (RT60) range from 0.1s to over 5s.
The mean and standard deviation of RT60s are 2.43$\pm$0.96s.
During the data curation process, noisy speech samples are manually filtered out via listening so that all speech labels are clean enough for accurate evaluations of the SE performance.
The improved label quality is also reflected by the \emph{all-positive} WADA-SNR values in Fig.~\ref{fig:wada_snr} (b) in contrast to (a).

Furthermore, to facilitate result analysis on the test data, Fig.~\ref{fig:hard_samples_test} additionally provides detailed \emph{tags} for each sample via manual labeling.
These tags as shown in the \textit{x}-axis are used to indicate the audio types, intelligibility levels, voice styles, speaker-related characteristics, degradation types, noise types, SNRs, and perceptual reverberation levels of each test sample.
A more detailed analysis will be presented in Section~\ref{ssec:bt_data}.

\vspace{-5pt}
\subsection{Blind test data}
\label{ssec:bt_data}
\vspace{-3pt}

The official blind test dataset consists of 500 simulated samples and 500 real recordings that can provide a more comprehensive evaluation of SE models' generalizability.
The simulated samples are prepared similarly to the non-blind test data in Section~\ref{ssec:nbt_data}, while the real recordings are collected from unseen sources.
Fig.~\ref{fig:wada_snr} (c) shows the estimated WADA-SNR distribution of the speech labels of the simulated samples.
Similar to the observation in Fig.~\ref{fig:wada_snr} (b), all estimated WADA-SNR values are positive in contrast to (a), indicating they are relatively cleaner than the training labels.
However, the highly unbalanced distribution of WADA-SNRs again implies the potential inaccuracy of the absolute WADA-SNR values as discussed in Section~\ref{ssec:tr_val_data}.

The hatched bars in Fig.~\ref{fig:hard_samples_test} provide detailed \emph{tags} of all degraded samples in the blind test set to facilitate data analysis.
By utilizing these tags, we are interested in answering the research question: \emph{Which attributes of speech samples are the most relevant in determining the difficulty of the SE task?}
To this end, we first divide the degraded samples in both test sets into two categories: hard samples and other samples.
A test sample is defined as ``hard'' if it is poorly enhanced by at least two participating teams, and a poorly enhanced  sample is determined by thresholding all evaluation metrics (to be introduced in Section~\ref{sec:metric}) with metric-wise weights\footnote{\url{https://github.com/urgent-challenge/urgent2024_analysis/blob/main/tagging/}.}.
Based on the above definitions, Fig.~\ref{fig:hard_samples_test} summarizes the tag occurrences of hard and other samples in both test sets.
For each tag in the \emph{x}-axis, two columns of bars are depicted, the left for non-blind test data and the right for blind test data.
By comparing the relative numbers of tag occurrences in hard and other samples, we have the following observations:
\begin{itemize}[wide]
	\item[1)] Modern SE systems tend to be robust against different voice styles, speaker genders, and different distortion types (e.g., bandwidth limitation, clipping). However, the following three acoustic conditions remain the most challenges for SE models to cope with: overlapped speech (denoted by the \texttt{speech\_overlap} and \texttt{unison} tags), strong wideband background noise and instantaneous noise, high reverberation.
    \item[2)] Samples with perceptually low intelligibility (often caused by strong degradations) are often hard to process. However, a reliable objective measure of speech sample difficulty is still unavailable. Conventional metrics such as SNR fail to correlate well with the difficulty, as indicated by the large percentage of hard samples in high-SNR (\textgreater 10 dB) blind test samples.
\end{itemize}

Inspired by these findings, several potential directions for improving existing SE pipelines can be summarized below:
\begin{itemize}
    \item[1)] Exploring better objective metrics that can reflect the speech sample difficulty more accurately, thus improving the design of SE corpora with balanced difficulty distribution.
    \item[2)] Designing more universal SE systems that can handle a wide variety of distortion types.
    \item[3)] Improving SE methods to tackle overlapped speech, strong noise and high reverberation.
\end{itemize}

{
\setlength{\tabcolsep}{3pt}
\begin{table*}
\setstretch{0.86}
\resizebox{\textwidth}{!}{
\begin{tabular}{cl|cc|cccccc|cc|cc|c|c}
\toprule
\multirow{2}{*}{\textbf{Rank}} & \multirow{2}{*}{\textbf{Team ID}} & \multicolumn{2}{c|}{\emph{Non-intrusive SE metrics}} & \multicolumn{6}{c|}{\emph{Intrusive SE metrics}} & \multicolumn{2}{c|}{\emph{Downstream-task-indep.}} & \multicolumn{2}{c|}{\emph{Downstream-task-dep.}} & \emph{Subjective} & \multirow{2}{*}{\textbf{\shortstack[c]{Overall\\ranking score}\,$\downarrow$}} \\
& & \textbf{DNSMOS}\,$\uparrow$ & \textbf{NISQA}\,$\uparrow$ & \textbf{PESQ}\,$\uparrow$ & \textbf{ESTOI}\,$\uparrow$ & \textbf{SDR}\,$\uparrow$ & \textbf{MCD}\,$\downarrow$ & \textbf{LSD}\,$\downarrow$ & \textbf{POLQA}\,$\uparrow$ & \textbf{SBS.}\,$\uparrow$ & \textbf{LPS}\,$\uparrow$ & \textbf{SpkSim}\,$\uparrow$ & \textbf{WAcc\,(\%)}\,$\uparrow$ & \textbf{MOS}\,$\uparrow$ \\ \midrule
1 & T1 & 3.06 (2) & 3.66 (3) & 2.65 (3) & 0.87 (2) & \textbf{14.58 (1)} & \textbf{3.04 (1)} & 2.92 (7) & 3.51 (2) & 0.84 (3) & 0.82 (4) & 0.80 (3) & 73.57 (2) & \textbf{3.52 (1)} & 2.43 \\
2 & T2 & 3.00 (6) & 3.59 (6) & \textbf{2.80 (1)} & \textbf{0.87 (1)} & 14.52 (2) & 3.15 (3) & 2.78 (4) & \textbf{3.69 (1)} & \textbf{0.85 (1)} & \textbf{0.83 (1)} & \textbf{0.82 (1)} & 72.91 (4) & 3.46 (3) & 2.90 \\
3a & T3a & 2.98 (9) & 3.44 (7) & 2.55 (6) & 0.85 (4) & 13.31 (4) & 3.33 (6) & 2.99 (9) & 3.34 (6) & 0.84 (5) & 0.83 (2) & 0.77 (7) & \textbf{74.03 (1)} & 3.44 (4) & 5.07 \\
3b & T3b & 2.95 (11) & 3.35 (11) & 2.66 (2) & 0.86 (3) & 13.54 (3) & 3.14 (2) & \textbf{2.70 (1)} & 3.45 (3) & 0.85 (2) & 0.83 (3) & 0.81 (2) & 73.10 (3) & 3.40 (7) & 5.07 \\
4 & T4 & 2.98 (8) & 3.37 (10) & 2.60 (4) & 0.85 (5) & 13.14 (5) & 3.21 (4) & 2.75 (3) & 3.43 (4) & 0.84 (4) & 0.81 (5) & 0.78 (5) & 71.67 (5) & 3.34 (10) & 6.53 \\
5 & T5 & 3.02 (4) & 3.60 (5) & 2.32 (9) & 0.82 (8) & 11.38 (10) & 3.34 (7) & 3.45 (14) & 3.16 (8) & 0.82 (9) & 0.78 (9) & 0.76 (8) & 67.96 (8) & 3.47 (2) & 6.57 \\
6 & T6 & 3.00 (7) & 3.35 (12) & 2.52 (8) & 0.84 (6) & 12.63 (6) & 3.32 (5) & 2.92 (8) & 3.31 (7) & 0.83 (6) & 0.80 (6) & 0.78 (6) & 70.13 (6) & 3.41 (6) & 6.83 \\
7 & T7 & 2.90 (16) & 3.38 (9) & 2.55 (5) & 0.83 (7) & 12.42 (7) & 3.61 (10) & 2.86 (5) & 3.36 (5) & 0.83 (7) & 0.79 (7) & 0.79 (4) & 69.19 (7) & 3.44 (5) & 7.30 \\
8 & T8 & 2.96 (10) & 3.15 (15) & 2.55 (7) & 0.80 (11) & 10.72 (11) & 3.83 (11) & 2.73 (2) & 3.15 (9) & 0.81 (11) & 0.75 (11) & 0.74 (11) & 66.15 (13) & 3.36 (9) & 10.60 \\
9 & T9 & 2.92 (14) & 3.42 (8) & 2.26 (11) & 0.80 (12) & 12.23 (8) & 4.12 (12) & 3.54 (16) & 3.04 (11) & 0.79 (12) & 0.74 (12) & 0.71 (12) & 67.03 (11) & 3.33 (11) & 11.43 \\
10 & T10 & 2.88 (18) & 3.17 (14) & 2.32 (10) & 0.81 (9) & 11.50 (9) & 3.46 (8) & 3.00 (10) & 3.06 (10) & 0.82 (10) & 0.77 (10) & 0.75 (9) & 67.45 (10) & 3.24 (13) & 11.57 \\
11 & T11 & 3.06 (3) & \textbf{3.94 (1)} & 1.88 (19) & 0.76 (15) & 7.49 (20) & 4.96 (20) & 4.76 (20) & 2.64 (17) & 0.75 (20) & 0.70 (17) & 0.58 (21) & 60.28 (19) & 3.39 (8) & 13.40 \\
12 & T12 & 2.92 (12) & 2.47 (21) & 2.14 (12) & 0.80 (10) & 9.73 (15) & 3.53 (9) & 3.36 (13) & 2.74 (14) & 0.83 (8) & 0.78 (8) & 0.75 (10) & 67.68 (9) & 2.87 (21) & 13.43 \\
13 & T13 & 2.89 (17) & 3.23 (13) & 2.03 (16) & 0.77 (14) & 10.43 (13) & 4.63 (16) & 3.83 (19) & 2.69 (15) & 0.77 (14) & 0.72 (14) & 0.67 (16) & 62.68 (15) & 3.32 (12) & 14.40 \\
14 & T14 & 2.88 (19) & 2.95 (18) & 2.13 (13) & 0.78 (13) & 10.62 (12) & 4.13 (13) & 3.24 (12) & 2.89 (12) & 0.77 (13) & 0.73 (13) & 0.70 (13) & 66.89 (12) & 3.06 (17) & 14.70 \\
\rowcolor[HTML]{EEEEEE}15 & Baseline & 2.83 (21) & 3.07 (17) & 2.07 (14) & 0.76 (16) & 10.13 (14) & 4.22 (15) & 3.09 (11) & 2.81 (13) & 0.77 (16) & 0.70 (16) & 0.70 (14) & 62.97 (14) & 3.12 (16) & 15.77 \\
16 & T16 & 2.92 (13) & 2.73 (19) & 2.04 (15) & 0.76 (17) & 9.47 (16) & 4.82 (19) & 3.55 (17) & 2.66 (16) & 0.77 (15) & 0.71 (15) & 0.67 (17) & 62.24 (16) & 2.95 (19) & 16.63 \\
17 & T17 & \textbf{3.26 (1)} & 3.83 (2) & 1.36 (22) & 0.60 (21) & 0.41 (22) & 6.27 (21) & 5.43 (21) & 1.74 (22) & 0.68 (21) & 0.56 (21) & 0.48 (23) & 40.73 (21) & 3.05 (18) & 16.80 \\
18 & T18 & 3.02 (5) & 3.61 (4) & 1.47 (21) & 0.51 (23) & -6.16 (23) & 8.44 (22) & 7.12 (23) & 1.93 (21) & 0.67 (22) & 0.53 (22) & 0.54 (22) & 32.08 (22) & 3.17 (15) & 17.13 \\
19 & T19 & 2.85 (20) & 3.12 (16) & 1.97 (18) & 0.74 (18) & 9.43 (17) & 4.65 (17) & 3.74 (18) & 2.59 (18) & 0.76 (18) & 0.69 (18) & 0.67 (18) & 60.28 (19) & 3.21 (14) & 17.23 \\
20 & T20 & 2.91 (15) & 2.55 (20) & 2.00 (17) & 0.73 (19) & 9.03 (19) & 4.18 (14) & 2.89 (6) & 2.57 (19) & 0.77 (17) & 0.68 (20) & 0.68 (15) & 60.64 (18) & 2.91 (20) & 17.63 \\
21 & T21 & 2.53 (22) & 2.39 (22) & 1.84 (20) & 0.73 (20) & 9.08 (18) & 4.74 (18) & 3.51 (15) & 2.47 (20) & 0.75 (19) & 0.68 (19) & 0.65 (19) & 59.95 (20) & 2.82 (22) & 20.20 \\
\rowcolor[HTML]{EEEEEE}22 & Noisy input & 1.70 (23) & 1.53 (23) & 1.26 (23) & 0.58 (22) & 0.98 (21) & 9.71 (23) & 5.46 (22) & 1.58 (23) & 0.59 (23) & 0.52 (23) & 0.64 (20) & 61.92 (17) & 1.88 (23) & 21.97 \\ \bottomrule
\end{tabular}
}
\caption{Final leaderboard of the URGENT 2024 Challenge on the blind test dataset. ``SBS.'' denotes the SpeechBERTScore. The non-intrusive SE metrics and WAcc\,(\%) are evaluated on the full blind test data. The intrusive SE metrics, downstream-task-independent metrics and SpkSim are evaluated on the simulated half of blind test samples. The subjective metric, i.e., MOS, is evaluated on a 300-sample subset as mentioned in Section~\ref{ssec:metric}.}
\label{tab:leaderboard}
\vspace{-1.5em}
\end{table*}
}

\vspace{-10pt}
\section{Metric Analysis}
\label{sec:metric}

In this section, we analyze and discuss the participants' results on the blind test dataset to evaluate and reflect on the design of the challenge's evaluation metrics.

\vspace{-5pt}
\subsection{Overview of evaluation metrics}
\label{ssec:metric}
\vspace{-5pt}

This section first provides an overview of the evaluation metrics used in the challenge, which are divided into five categories for a comprehensive evaluation:
\begin{itemize}[wide,labelwidth=!,labelindent=0pt]
	\item[1)] Non-intrusive SE metrics: DNSMOS and NISQA~\cite{NISQA-Mittag2021}.
	\item[2)] Intrusive SE metrics: POLQA~\cite{POLQA-Beerends2013}, PESQ~\cite{PESQ-Rix2001}, ESTOI~\cite{ESTOI-Jensen2016}, signal-to-distortion ratio (SDR)~\cite{SDR-Vincent2006}, mel cepstral distortion (MCD)~\cite{MCD-Kubichek1993}, log-spectral distance (LSD)~\cite{LSD-Gray1976}.
	\item[3)] Downstream-task-independent metrics: Levenshtein phoneme similarity (LPS)~\cite{Evaluation-Pirklbauer2023} and SpeechBERTScore~\cite{SpeechBERTScore-Saeki2024}.
	\item[4)] Downstream-task-dependent metrics: speaker similarity (SpkSim) and word accuracy (WAcc).
	\item[5)] Subjective metric: mean opinion score (MOS).
\end{itemize}
Compared to the initial design in~\cite{URGENT-Zhang2024}, two new metrics (i.e., POLQA and MOS) are added to better reflect the quality of the enhanced audio, and the same ranking rule as in~\cite{URGENT-Zhang2024} is adopted to obtain the final results.
Note that the MOS evaluation was conducted via ITU-T P.808~\cite{P.808,Open-Naderi2020} tests on a 300-sample subset (half simulated and half real-recorded) of the blind test data due to limited time and budget.
The subset was carefully selected to preserve the full test set distribution in terms of domains, difficulty and degradations. 
The subjective listening test included the noisy samples, baseline results, and enhanced samples from all 21 participants' submissions in the final blind test phase.
Each sample was evaluated by eight different listeners vetted through a strict qualification procedure.

The final results are summarized in Table~\ref{tab:leaderboard}, where the overall ranking score in the last column is  the weighted sum of all metrics' individual rankings~\cite{URGENT-Zhang2024}.
We can see that the top-ranking teams generally excel in all metrics, while the individual ranking of each metric can differ from the overall ranking.
This highlights the necessity of combining various metrics to measure the audio quality comprehensively.
On the other hand, the remaining teams often show mixed rankings among different metrics.
For example, teams \texttt{T17} and \texttt{T18} show excellent performance on non-intrusive SE metrics (i.e., DNSMOS and NISQA), with much worse performance on all other metrics including MOS.
This interesting phenomenon reveals the importance of using multifaceted metrics than purely non-intrusive objective metrics, which sometimes occur in SE studies using real data~\cite{RaD_Net2-Liu2024,Personalized-Prnamaa2024,PESQetarian-deOliveira2024}.

\begin{figure}[t]
  \centering
  \includegraphics[width=0.9\columnwidth]{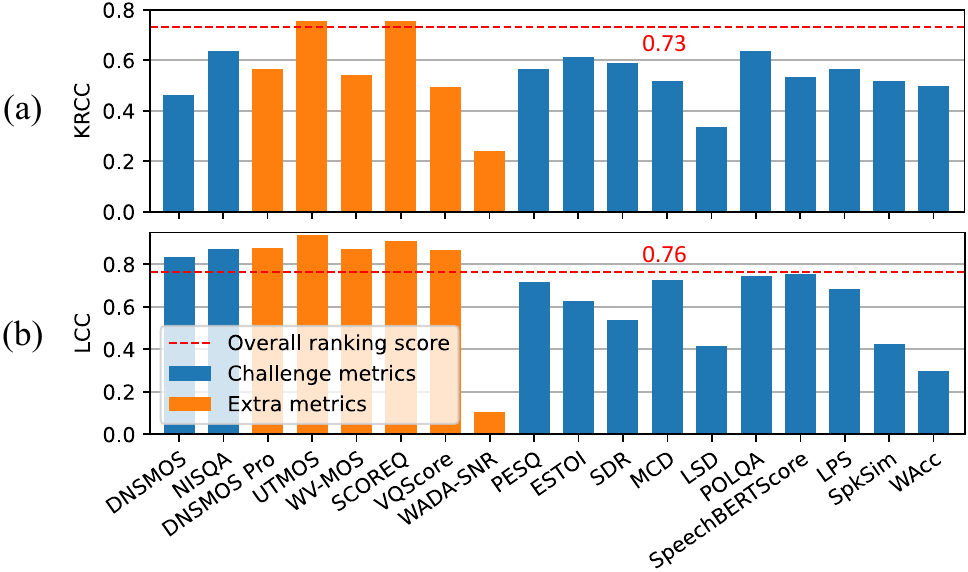}
  \caption{Correlations (KRCC and LCC) between MOS and other objective metrics on the blind test data.}
  
  \label{fig:mos_corr}
\end{figure}

\vspace{-8pt}
\subsection{Perceptual correlation of different objective metrics}
\label{ssec:corr}
\vspace{-5pt}

Following the analysis above, we further investigate how each objective metric reflects the actual perceptual quality.
Thanks to the availability of the subjective metric in Table~\ref{tab:leaderboard}, we can easily calculate the correlation between each objective metric and human judgments (i.e., MOS).
Furthermore, we include 6 extra non-intrusive metrics for comparison: DNSMOS Pro~\cite{DNSMOSPro-Cumlin2024}, UTMOS~\cite{UTMOS-Saeki2022}, WV-MOS~\cite{WVMOS-Andreev2023}, SCOREQ~\cite{SCOREQ-Ragano2024}, VQScore~\cite{VQScore-Fu2024}, and WADA-SNR~\cite{WADASNR-Kim2008}.
As shown in Fig.~\ref{fig:mos_corr} (a), we adopt the Kendall rank correlation coefficient (KRCC)~\cite{KRCC-Kendall1938}
to evaluate the relevance between the ranks of MOS and other individual objective metrics.
Fig.~\ref{fig:mos_corr} (b) further presents the linear correlation coefficient (LCC)~\cite{LCC-Pearson1920} between the MOS score and the score of each objective metric.
The KRCC between the MOS rank and the overall rank is 0.73, showing strong consistency between the MOS and the overall ranking score.
Comparing individual metrics, we have the following findings:
\begin{itemize}[wide,labelwidth=!,labelindent=0pt]
	\item[1)] Among all objective metrics, UTMOS and SCOREQ demonstrate the highest correlations with MOS.
	In contrast, WADA-SNR shows the lowest correlations with MOS, suggesting that it is unsuitable for measuring enhanced audio quality.
	\item[2)] Among all challenge metrics, LSD and DNSMOS lag behind others on KRCC, indicating that they may not accurately reflect the overall SE performance when used alone.
    Meanwhile, the relatively high LCC value of DNSMOS implies that the rank mismatch between MOS and DNSMOS can be large while their numerical values are close (e.g., rows 12 and 16 in Table~\ref{tab:leaderboard}).
	In comparison, other non-intrusive metrics such as NISQA and DNSMOS Pro tend to be more consistent with MOS in terms of both ranks and values.
	\item[3)] SpkSim and WAcc show relatively high KRCCs with much lower LCCs, suggesting that downstream-task-dependent metric values are likely to correlate non-linearly with the audio perceptual quality.
\end{itemize}

Based on the above analysis, the best practice of SE evaluation can be summarized as follows:
\begin{itemize}
    \item[1)] Adopting advanced non-intrusive metrics (e.g., UTMOS and SCOREQ) in lieu of\,/\,along with traditional ones (e.g., DNSMOS) for more accurate SE performance assessment.
    \item[2)] Utilizing multifaceted metrics for comprehensive SE evaluations rather than using a single category of metrics, especially purely non-intrusive ones.
\end{itemize}

\vspace{-8pt}
\section{Conclusion}
\label{sec:conclusion}
\vspace{-3pt}

In this paper, we have presented an in-depth analysis of the recent URGENT 2024 Challenge from data and metric perspectives.
By looking into the distributions of various data attributes (e.g., sampling frequency, estimated SNR, tag occurrence), we have revealed several missing or overlooked issues on data preparation in the literature.
Building on the comprehensive evaluation metric design of the challenge, we further examined the effectiveness of various metrics, including recently proposed ones that were not originally used during the challenge.
Based on the above analysis, we provided insights into the effectiveness of various metrics, which we hope will inspire improved designs or combinations of evaluation metrics in the future.
In our future work, we plan to delve deeper into other aspects of the challenge outcomes, such as model architectures and training strategies.

\ifinterspeechfinal
\section{Acknowledgment}
\vspace{-4pt}
The leaderboard evaluation has been supported by the PSC Bridges2 system via ACCESS allocation CIS210014, supported by National Science Foundation grants \#2138259, \#2138286, \#2138307, \#2137603, and \#2138296.
The subjective listening test was funded and executed by Technische Universität Braunschweig, strictly following ITU-T Recommendation P.808 \cite{P.808, Open-Naderi2020}.
It has been implemented on the Amazon Mechanical Turk (MTurk) platform, delivering absolute category rating (ACR) MOS scores.
This work was supported in part by China STI 2030-Major Projects under Grant No. 2021ZD0201500, in part by China NSFC projects under Grants 62122050 and 62071288, and in part by Shanghai Municipal Science and Technology Commission Project under Grant 2021SHZDZX0102.
\fi
\vspace{-6pt}
\bibliographystyle{IEEEtran}
\bibliography{mybib}

\end{document}